\documentclass[times,twocolumn,final]{elsarticle}

\usepackage{medima}
\usepackage{framed,multirow}

\usepackage{amssymb}
\usepackage{latexsym}

\usepackage{url}
\usepackage{color,colortbl,booktabs}
\usepackage[table,x11names,dvipsnames,table]{xcolor}
\definecolor{lightgray}{gray}{0.95}

\usepackage{hyperref}
\hypersetup{
    colorlinks=true,
    linkcolor=blue,
}

\usepackage{float,tablefootnote,threeparttable,wrapfig}

\usepackage{booktabs}
\usepackage{subcaption}

\usepackage{lipsum}
\usepackage{color}
\usepackage[ruled,vlined]{algorithm2e}
\usepackage{diagbox} 
\usepackage{tabu}
\usepackage{arydshln}

\newcommand{\revision}[1]{\textcolor{black}{#1}} 

\usepackage{xcolor}
\definecolor{ForestGreen}{rgb}{0.13, 0.55, 0.13}
\usepackage{pifont}
\newcommand{\cmark}{{\color{ForestGreen} \ding{51}}}%
\newcommand{\xmark}{{\color{red} \ding{55}}}%

\usepackage{amsmath} 
\usepackage{amssymb}  
\usepackage{diagbox} 
\usepackage{tabu}
\usepackage{algpseudocode}
\usepackage{epsfig}
\usepackage{hyperref}
\usepackage{booktabs}  
\usepackage{multirow}  
\usepackage{array}

\journal{Medical Image Analysis}

\begin{document}

\verso{Z. Jiang \textit{et~al.}}

\begin{frontmatter}

\title{DefCor-Net: Physics-Aware Ultrasound Deformation Correction}

\author[1]{Zhongliang \snm{Jiang}*\corref{cor1}}
\cortext[cor1]{The first two authors contributed equally to this work. \\
Corresponding author at: Technical University of Munich, Faculty of informatics -- I16, Boltzmannstr. 3, 85748 Garching bei M{\"u}nchen. \\
This work involved human subjects in its research. Approval of all ethical and experimental procedures and protocols was granted by Institutional Review Board, No. 2022-87-S-KK, Declaration of Helsinki.
}
\ead{zl.jiang@tum.de}

\author[1]{Yue \snm{Zhou}*}

\author[2]{Dongliang \snm{Cao}}

\author[1,3]{Nassir \snm{Navab}}

\address[1]{Computer Aided Medical Procedures, Technical University of Munich, Munich, Germany}
\address[2]{University of Bonn, North Rhine-Westphalia, Germany}
\address[3]{Computer Aided Medical Procedures, Johns Hopkins University, Baltimore, MD, USA}

\received{XX June 2022}
\finalform{xx Month 2022}
\accepted{xx Month 2022}
\availableonline{xx Month 2022}

\begin{abstract}
The recovery of morphologically accurate anatomical images from deformed ones is challenging in ultrasound (US) image acquisition, but crucial to accurate and consistent diagnosis, particularly in the emerging field of computer-assisted diagnosis. This article presents a novel anatomy-aware deformation correction approach based on a coarse-to-fine, multi-scale deep neural network (DefCor-Net). To achieve pixel-wise performance, DefCor-Net incorporates biomedical knowledge by estimating pixel-wise stiffness online using a U-shaped feature extractor. 
The deformation field is then computed using polynomial regression by integrating the measured force applied by the US probe.
Based on real-time estimation of pixel-by-pixel tissue properties, the learning-based approach enables the potential for anatomy-aware deformation correction. To demonstrate the effectiveness of the proposed DefCor-Net, images recorded at multiple locations on forearms and upper arms of six volunteers are used to train and validate DefCor-Net. The results demonstrate that DefCor-Net can significantly improve the accuracy of deformation correction to recover the original geometry (Dice Coefficient: from $14.3\pm20.9$ to $82.6\pm12.1$ when the force is $6N$). \\
\textbf{Code:} \url{https://github.com/KarolineZhy/DefCorNet}.
\end{abstract}

\end{frontmatter}


\section{Introduction}
\label{introduction}

\begin{figure}[ht!]
\centering
\includegraphics[width=0.45\textwidth]{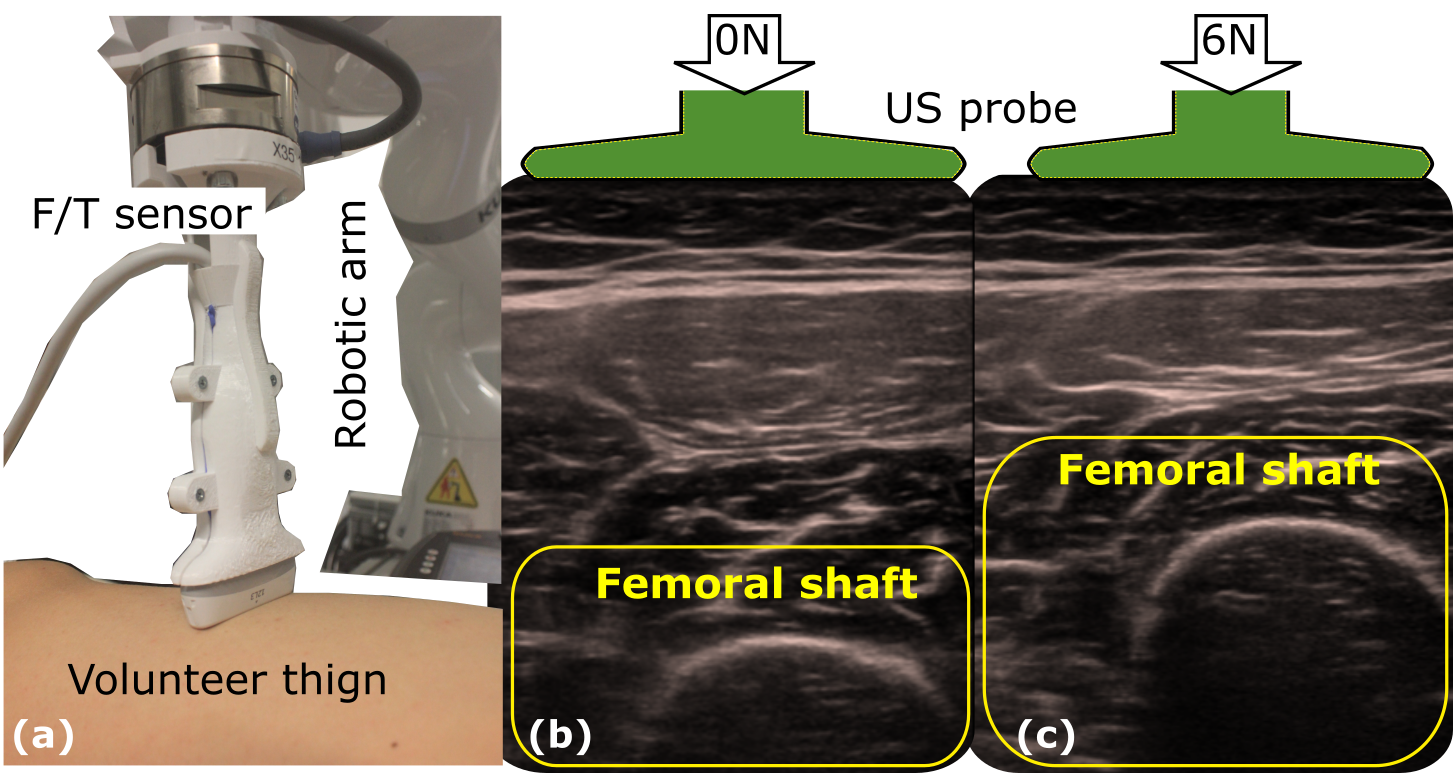}
\caption{Illustration of the effects of pressure-induced deformation. (a) robotic arm with a force/torque (F/T) sensor. (b) and (c) are the resulting B-mode images of the femoral shaft recorded when the contact force is $0~N$ and $6~N$, respectively.
}
\label{Fig_defromation_background}
\end{figure}

\par
The recovery of accurate anatomical images from distorted ones is a fundamental problem in medical imaging, which contributes to the achievement of reliable diagnoses and biometric measurements in clinical practices. Regarding medical ultrasound (US), it has been widely used for examining internal organs due to its merits of low cost, high accessibility, and lack of radiation. To optimize the acoustic coupling performance, a certain pressure needs to be applied due to the inherited characteristic of US modality. Inconsistent pressure will result in distorted images with varying anatomy geometry for soft tissue, such as superficial veins~\citep{jiang2021deformation}, preventing the achievement of consistent diagnosis results among different clinicians. In contrast to experienced sonographers who are trained and accustomed to making diagnoses based on distorted images, image distortion will have a significant impact on the development of autonomous diagnosis systems that aim to produce standardized diagnoses~\citep{wang2020automatic}.
In addition to providing the precise geometries of soft tissues, the recovered images also reveal the precise locations of high-stiffness anatomies, e.g., bone (see Fig.~\ref{Fig_defromation_background}). This is crucial for modern image-guided orthopedic surgery, which requires accurate registration between intra-operative US scans (bone surface) and pre-operative MRI or CT scans to transfer the planned trajectory~\citep{jiang2023thoracic, jiang2023skeleton, wein2015automatic, salehi2017precise}.

\subsection{Robotic US Imaging}
\par
Due to their stability and accuracy, robotic US systems (RUSS) are viewed as a promising solution for obtaining high-quality images. In contrast to free-hand US imaging, the US acquisition parameters (contact force and probe orientation) can be accurately maneuvered in RUSS~\citep{gilbertson2015force}. To precisely control the probe orientation, Jiang~\emph{et al.} estimated the normal direction of unknown constraint surfaces using both force cues and US confidence map~\citep{jiang2020automatic}. The US confidence map provides pixel-wise quality measurements (signal strength) based on the physics of sound propagation model~\citep{karamalis2012ultrasound}. To enable the autonomous adjustments of probe orientation for both linear and convex probes, Jiang~\emph{et al.} developed a mechanical model-based algorithm relying on accurate measurements of contact force~\citep{jiang2020automatic_tie}. In addition, Ma~\emph{et al.} employed a depth camera to estimate the normal direction of the chest surface for screening of the lung~\citep{ma2021autonomous}. Although the camera-based approach is less accurate than the force-based method, the computation is performed more quickly. In addition to optimizing probe orientation, a number of force control schemes have been proposed for RUSS~\citep{gilbertson2015force, zettinig20173d, pierrot1999hippocrate}. Such approaches can maintain a given force between the probe and the contact surface, which can alleviate the inter-variation of pressure-induced deformation between the images acquired along a scanning path. 
Nevertheless, due to variations in the physical properties of human tissues (e.g., bone and muscle), image deformation for various anatomies in individual B-mode images remains inconsistent.
To maintain the focus of this article, we refer readers to two recent survey articles for more detailed developments in the field of RUSS~\citep{jiang2023robotic, li2021overview}.

\subsection{US Imaging Deformation Correction}
\par
To recover uncompressed images from the deformed ones, Treece~\emph{et al.} combined non-rigid image-based registration and external position sensing~\citep{treece2002correction}. This pioneering study considered both global accuracy and local accuracy based on the position sensor measurements and correction of the pressure-induced errors, respectively. But this approach was developed for axial deformation. To further consider the deformation in the lateral direction, Burcher~\emph{et al.} built an elastic model using the finite element method (FEM), where the tissue parameters were manually assigned~\citep{burcher2001deformation}. Similarly, Flach~\emph{et al.} developed a correction approach in FEM based on the assumption that the tissues are homogeneous~\citep{flack2016model}. To address this limitation, Sun~\emph{et al.} proposed a novel trajectory-based method to recover compression-free images using an empirical regressive model with respect to the contact force~\citep{sun2010trajectory}. To obtain the displacement trajectories of individual pixels under varying contact forces, a template-based image-flow technique was applied to radio frequency data. Similarly, Virga~\emph{et al.} used $4$th-order polynomial models to depict specific pixel displacement by manually annotating $5$ feature points in all frames obtained with different forces; and then, a novel graph-based inpainting approach was employed to compute 2D deformation fields for unsampled pixels~\citep{virga2018use}. Although some promising results were reported on in-vivo human tissues, the lengthy computational time ($186~s$ per inpainting procedure) hinders further clinical translation.

\par
In contrast to the aforementioned deformation correction approaches that directly link the deformation to the contact force, another folder of methods also takes tissue stiffness into consideration. Dahmani~\emph{et al.} first estimated the relative mechanical parameters of involved tissues based on the images themselves with a reference baseline; then, a model-based correction approach was developed to recover deformed images~\citep{dahmani2017model}. Regarding the deformation recovery task, their method demonstrated better performance than the image-based method~\citep{sun2010trajectory}. But the validations were only carried out on simplified synthetic US images. In addition, Jiang~\emph{et al.} presented a stiffness-based deformation correction method, incorporating image pixel displacements, contact forces, and nonlinear tissue stiffness~\citep{jiang2021deformation}. In their study, the tissue stiffness was characterized using the probe displacements and contact forces recorded during robotic palpation, while the pixel displacements were computed using the LukasKanade algorithm~\citep{lucas1981iterative}. To demonstrate the propagation capability of the proposed approach for various patients, the optimized deformation regression model computed at sampled positions on a stiff phantom can be successfully propagated to unsampled positions on the same stiff phantom and on a commercial soft phantom, respectively, by only substituting the estimated tissue stiffness. Nevertheless, limited by the number of coefficients of the polynomial model, the accuracy for individual pixels is sacrificed to ensure overall optimal performance for all pixels. Consequently, the performance of this approach is spatially location-dependent. In addition, the sparse optical flow technique hinders the achievement of dense deformation correction.

\subsection{Dense Flow Estimation}
\par
The deformation correction process is often considered an inverse problem, where the deformation field is computed first and further used for recovery. The sparse optical flow technique is often used to extract pixel displacements for computing the deformation field~\citep{jiang2021deformation, sun2010trajectory, virga2018use}. Nevertheless, the sparse optical flow cannot generate a pixel-by-pixel deformation field. Thanks to the booming of deep learning, the state-of-the-art dense flow estimation approaches have achieved phenomenal success, e.g., FlowNet~\citep{dosovitskiy2015flownet,ilg2017flownet}, PWC-Net~\citep{sun2018pwc} and Recurrent All-Pairs Field Transforms (RAFT)~\citep{teed2020raft}. Dosovitskiy~\emph{et al.} first cast optical flow estimation as a learning problem and solve it based on a standard Convolutional neural networks (CNN) architecture~\citep{dosovitskiy2015flownet}. It was reported that FlowNet could accurately estimate the flow at a frame rate of $5$ to $10~fps$. To further improve time efficiency and flow quality, Ilg~\emph{et al.} developed FlowNet 2.0 by introducing a stacked architecture by warping the second image with intermediate optical flow~\citep{ilg2017flownet}. To simultaneously increase the accuracy and decrease the size of a CNN model for optical flow, Sun~\emph{et al.} presented PWC-Net, a coarse-to-fine network architecture with pyramid, warping, and a cost volume~\citep{sun2018pwc}. The results demonstrated that PWC-Net is approximately $17$ times smaller in size, and two times faster than FlowNet 2.0 for inference. Instead of the coarse-to-fine architecture, Teed~\emph{et al.} proposed RAFT, which maintains and updates a single fixed flow field at high resolution~\citep{teed2020raft}. To reliably predict the flow field even when the displacement is large, a recurrent GRU-based operator is employed. In addition to its superior estimation accuracy relative to the aforementioned methods, RAFT has also been reported to possess strong cross-dataset generalization~\citep{teed2020raft}.

\subsection{Ultrasound Elastography}
\par
US elastography~\citep{sigrist2017ultrasound} is an evolutionary method that uses US waves to assess the mechanical properties of tissues, such as stiffness. Particularly for strain elastography (also called static elastography), displacement estimation between two frames is one of the most crucial steps for high-quality elastography~\citep{rivaz2014ultrasound,hashemi2018assessment,rivaz2008ultrasound}. From this point of view, the deformation correction procedure is an inverse problem of strain elastography. To compute the pixel-wise displacement, Rivaz~\emph{et al.} minimized cost functions incorporating the similarity of radio–frequency (RF) data intensity and displacement continuity of two frames~\citep{rivaz2010real}. To improve computational effectiveness, Ashikuzzaman~\emph{et al.} considered the RF data from three consecutive frames to ensure that both spatial and temporal estimations are data-adaptive~\citep{ashikuzzaman2019global}. To improve the quality of elastography, Guo~\emph{et al.} employed a partial differential equation (PDE)–based regularization algorithm to iteratively reduce noise contained in the displacement estimations~\citep{guo2015pde}. Considering the establishment of elastography is often time-consuming, Boctor~\emph{et al.} proposed a rapid approach to segment stiff lesions for monitoring ablative therapy based on the tissue deformation map~\citep{boctor2006ultrasound}.

\par
In addition to model-based techniques for estimating mechanical parameters from force-displacement measurements, the development of neural networks offers an additional folder of model-free solutions. Hoerig~\emph{et al.} presented a data-driven approach to compute elasticity images using artificial neural networks~\citep{hoerig2018data}. This approach does not require any prior assumptions about the underlying constitutive model, but it does require internal structure information (discrete or continuous material property distributions). Tehrani~\emph{et al.} computed the displacement by training an optical flow network first using computer vision datasets and followed by a fine-tuning process to bridge the gap between the natural images and the RF data~\citep{tehrani2022bi}. 
To enhance the accuracy of lateral strain estimation, Tehrani~\emph{et al.} proposed a novel framework using physically inspired constraint (effective Poisson’s ratio) for unsupervised regularized elastography~\citep{tehrani2022lateral}. In order to obtain dense elastography, RF data is often used as the input rather than the B-mode images. Yet, RF data is not accessible for most of the commercial US machines used in clinical practices. Due to the fact that the estimated stiffness of various tissues is only relative, it cannot be directly adapted to recover the force-induced deformation for a single image, i.e., individual images with force information.

\subsection{Proposed Approach}
\par
To recover zero-compression images from deformed ones, we present a novel anatomy-aware deformation correction approach using a deep neural network (DefCor-Net) for achieving accurate anatomical US imaging without reference images. To the best of the authors' knowledge, this is the first work introducing a learning-based approach to correct the pressure-induced US deformation with pixel-wise and real-time performance. The DefCor-Net mainly consists of a stiffness estimation module (SEM) and a deformation field estimation module (DFM). The SEM first computes the local stiffness online from an input image using a U-shape network unit~\citep{ronneberger2015u}. Then, the patient-specific tissue stiffness computed based on robotic palpation is used to update the computed local stiffness map, thereby enabling the capability to generalize across patients. In addition, the DFM is designed to predict pixel-wise deformation fields based on the physics of the interaction between the current force and the computed stiffness map. The DefCor-Net is implemented in a coarse-to-fine manner to capture both low-resolution and high-resolution characteristics, thereby facilitating the computation of the pixel-wise deformation field, even for relatively large deformations. The DefCor-Net is trained in an end-to-end fashion, where the ground truth of the deformation field is represented by the dense optical flow map computed using RAFT~\citep{teed2020raft}. Since RAFT can only estimate the displacement based on two frames, they cannot be used to correct individual deformed images without reference frames. Contrary to conventional polynomial model-based methods~\citep{sun2010trajectory,jiang2021deformation,dahmani2017model}, DefCor-Net has a large number of trainable parameters, i.e., $7,782,936$ in total, which enables the achievement of pixel-wise correction results due to the dense stiffness map. The experiments were carried out on the in-vivo data recorded from six volunteers, including multiple sampling locations on both forearms and upper arms.

\section{Material and Data Preparation}

\subsection{Hardware Setup}
\par
Our robotic US system consists of a lightweight robot, a US device, and an accurate F/T sensor. The F/T sensor was directly mounted on the robotic flange in order to precisely measure the contact force. The US probe was attached on the other side of the F/T sensor using a custom-designed holder.   

\subsubsection{Robotic Manipulator}
\par
In this study, a redundant robotic arm (KUKA LBR iiwa 7 R800, KUKA Roboter GmbH, Germany) with seven joints was used. In each joint, a torque sensor is incorporated, allowing the implementation of compliant force control to address safety concerns and acoustic coupling issues. To control the robotic arm, an open-source interface\footnote{https://github.com/IFL-CAMP/iiwa\_stack} developed in our lab was used. This interface allows direct communication between the low-level Sunrise applications and the Robot Operating System (ROS) framework. To guarantee real-time performance, the robotic status and the control commands from user applications were updated in $500~Hz$ and $100~Hz$, respectively.  

\subsubsection{Tactile Sensing}
\par
To accurately measure the dynamic contact force, an external F/T sensor (Gamma, ATI IndustrialAutomation, USA) was used. The measured force can be accessed using a data acquisition device (FTD-DAQ-USB6361, National
Instruments, USA). To synchronize the force information and US images, the force was published to the ROS master with precise timestamps at a rate of $100~Hz$.

\subsubsection{Ultrasound System}
\par
In this study, B-mode images were acquired using an ACUSON Juniper US machine (Siemens Healthineers, Germany) with a linear probe (12L3, Siemens Healthineers, Germany). The B-mode images were updated in $58~fps$. To access the B-mode images from the US machine, a frame grabber (Epiphan DVI2USB 3.0, Epiphan Vision, Canada) was used to take screenshots in $30~fps$. To synchronize the force information and US images, the US images stream was published to the ROS master in real-time. Then the images and force data can be synchronized using OpenIGTLink\footnote{http://openigtlink.org/}, where the time tolerance was $12~ms$. The US images were recorded using the default configuration for bone. The detailed parameters are listed as follows: imaging depth: $45~mm$, focus: $20~mm$, frequency: $6.7~MHz$, and brightness: $78~dB$.

\subsection{Compliant Force Control Architecture}
\par
In order to estimate the patient-specific mechanical properties and record paired data (contact force, probe pose, and US images), ``robotic palpation" was performed by increasing and decreasing the applied force. To this end, a compliant force controller with an input of desired force was used~\citep{hennersperger2016towards}. Compared to a position controller, the compliant mode can provide a safer interaction avoiding excessive force. The Cartesian compliant controller is defined as Eq.~(\ref{eq_impedance_law}).

\begin{equation}\label{eq_impedance_law}
\tau = \textbf{J}^{T}[\textbf{F}_d + \textbf{K}_m e + \textbf{D} \dot{e} + \textbf{M} \Ddot{e}]
\end{equation}
where $\tau \in \textbf{R}^{7\times1}$ is the required joint torque to realize the compliant force controller, $J^{T}\in \textbf{R}^{7\times6}$ is the transposed Jacobian matrix, $e\in \textbf{R}^{6\times1} = (x_d - x_c)$ is the pose error (position and orientation) between the current pose $x_c$ and the desired pose $x_d$ in Cartesian space, $\textbf{F}_d\in \textbf{R}^{6\times1}$ is the desired F/T applied at the tool center point (TCP), $\textbf{K}_m \in \textbf{R}^{6\times6}$, $\textbf{D}\in \textbf{R}^{6\times6}$ and $\textbf{M}\in \textbf{R}^{6\times6}$ are diagonal matrices of stiffness, damping and inertia values in $6$ degree of freedom (DoF), respectively. According to~\citep{hennersperger2016towards}, the stiffness is suggested to be in the range of $[125, 500] N/m$ for different tissues. In addition, $25~N$ was set as the maximum contact force in the low-level safety configuration in the robotic cabinet, which will be triggered once a force larger than $25~N$ occurs.


\subsection{Robot-Assisted Data Collection}

\par
To record in-vivo US images, six healthy volunteers were employed. For each volunteer, we evenly selected four sampling points on the forearm from the elbow joint toward the wrist joint in $150~mm$ distance. In addition, two sampling points were selected on the upper arm with a distance of $40~mm$. Each robotic palpation consists of one cycle of pressing and releasing, in which the applied force was gradually increased from zero to $15~N$ and then decreased to zero using the compliant force controller. Each palpation took around $30~s$, which will produce approximately $900$ sequential B-mode images. At each sampling location, robotic palpation were carried out twice to collect in-vivo data.
In the end, we have $72$ image sets on both forearms and upper arms from six volunteers. To train the DefCor-Net, $51$ ($~70\%$) and $8$ ($~10\%$) image sets were used for training and validation, respectively. The remaining $13$ unseen image sets were used for testing. 

\par
Based on the recorded position and measured force at the probe tip, the global stiffness of human tissue can be estimated (see Section~\ref{subsection: stiffness}). 
The average stiffness values ($\pm$SD) of the forearm and upper arm over sampling points were $1.80\pm0.48~N/mm$ and $0.78\pm0.03~N/mm$, respectively. The maximum compression levels that happened at the contact surface for the upper arm and forearm were $12.5\pm1.14~mm$ and $7.1\pm0.8~mm$, respectively. Both results indicate that the mechanical properties of the forearm and upper arm are significantly different.
During the process of data acquisition, volunteers were instructed to extend their arms naturally on a flat table.


\begin{figure*}[h!]
    \centering
    \includegraphics[width=0.75\textwidth]{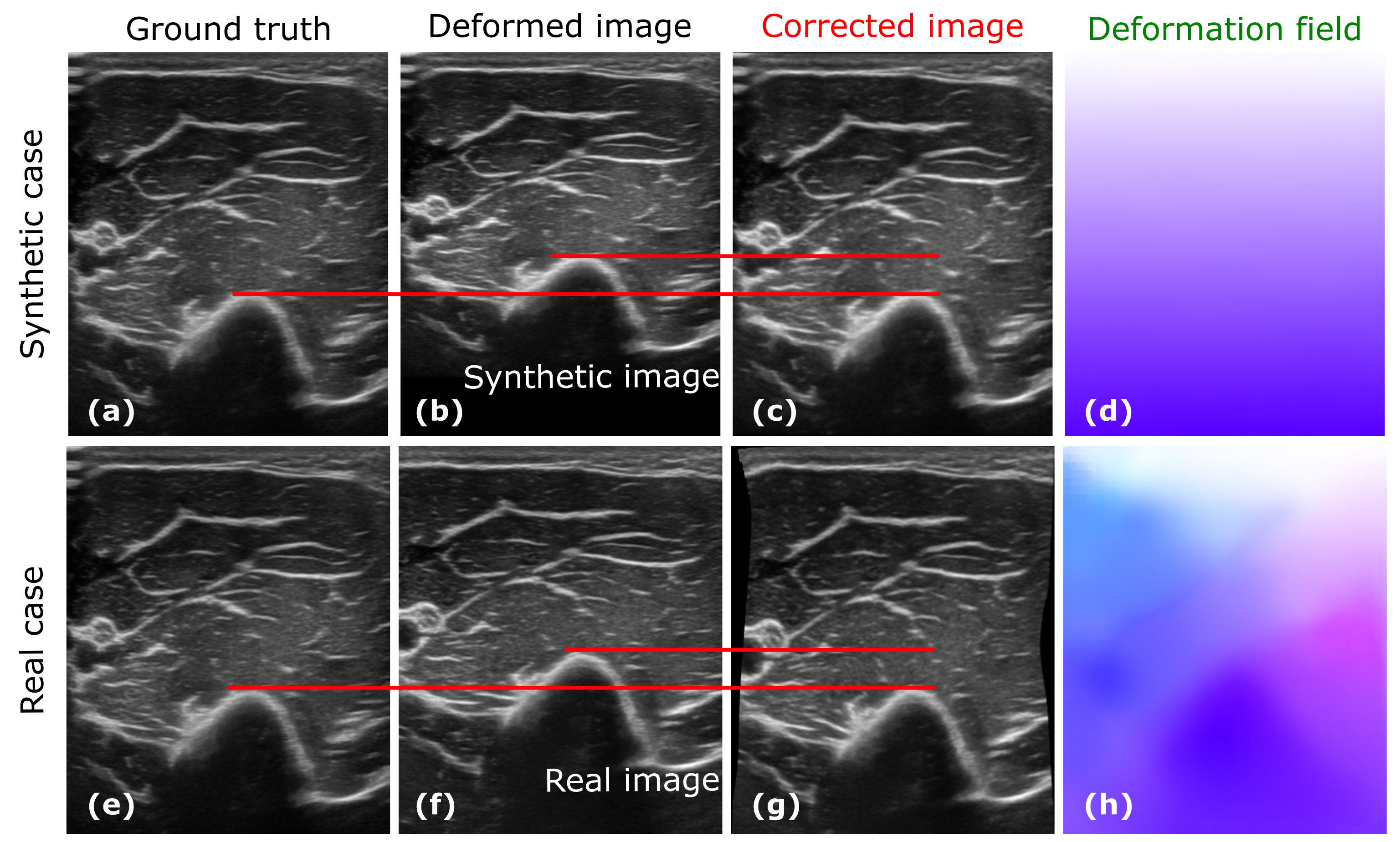}
    \caption{Illustration of the computed deformation field and the corrected results. The top and bottom rows represent the representative results on synthetic data and real data acquired when the contact force is $7.5~N$. For the synthetic case, the deformed images is generated by evenly moving the individual pixels with different displacements, where the top row is moved zero, and the bottom row is moved $70$ pixels. The deformation fields are computed based on the deformed images and ground truth images. The corrected images in both cases are close to the ground truth. The Normalized Cross Correlation (NCC) similarity between two images is (a,b)=$0.31$ (c,b)=$0.99$; (e,f)=$0.34$, (g,f)=$0.86$.
    }
    \label{fig_ground_truth_raft}
\end{figure*}

\subsection{Optical Flow-Based Deformation Field Generation}~\label{sec:gt_raft}
\par
The estimation of the deformation field is the key to the deformation correction procedure. Benefiting from the boom of learning-based dense optical flow estimation technologies, this study explores the feasibility of using the computed dense flow map to represent the deformation field between two images. Since the deformation field between a deformed image and its corresponding uncompressed image can be estimated offline, the accuracy of flow estimation becomes the most crucial issue. Due to the superiority in terms of accuracy and generalization for different datasets, RAFT with a well-trained model (Sintel)\footnote{https://github.com/princeton-vl/RAFT} was employed to compute the dense optical flow field representing the deformation field. Considering RAFT is originally designed to estimate image flow based on two consecutive images, an intermediate frame $\mathbb{I}_{m}$ between the uncompressed image $\mathbb{I}_{1}$ and the current image $\mathbb{I}_{2}$ is used to guarantee the flow estimation accuracy by avoiding too large displacement. The final flow $f_{2,1}$ between $\mathbb{I}_{1}$ and $\mathbb{I}_2$ is calculated as Eq.~(\ref{eq_flow_combine}). 

\begin{equation}~\label{eq_flow_combine}
    f_{1,2} = f_{1,m} + warp(f_{m,2}, f_{1,m})
\end{equation}
where $f_{i,j}$ represents the image flow estimated from image $\mathbb{I}_{i}$ to $\mathbb{I}_{j}$ and $warp(\mathbb{I}_{j}, f_{i,j}) \rightarrow \mathbb{I}_{i}$ represents the warping operation to recover deformed image $\mathbb{I}_{j}$ using known flow map $f_{i,j}$. For each pixel location $P_k$ in $\mathbb{I}_{i}$ and its corresponding pixel in $\mathbb{I}_{j}$ are satisfied the following equation. 

\begin{equation}~\label{eq_warping}
\begin{split}
    &c^j[P_k+ f_{i,j}(P_k)] = c^i(P_k) \\
    &s.t.~~~ [0,0]^T\leq P_k + f_{i,j}(P_k)\leq [W, H]^T
\end{split}
\end{equation}
where $c^i$ and $c^j$ are the intensity values of a pixel position on $\mathbb{I}_{i}$ and $\mathbb{I}_{j}$, respectively. $W$ and $H$ are the width and height of images in terms of the pixel. The bilinear interpolation is used to implement the warping operation as~\citep{teed2020raft}. To intuitively demonstrate the accuracy of the generated deformation field and the correction performance of deformed images, two representative results from synthetic and real deformed images are depicted in Fig~\ref{fig_ground_truth_raft}.

\par
Due to the force-induced compression, anatomical structures on US images are moved towards the top of the image. To validate whether the RAFT can correctly compute the deformation field, a synthetic image was generated by evenly moving pixel lines (horizontal) at different distances. In our case, the top line (contact surface) was kept constant, while the bottom line was moved $70$ pixels ($18.2\%$ of the image height) in the vertical direction. The computed deformation field is demonstrated in Fig.~\ref{fig_ground_truth_raft}~(d), which conforms the expectation. No displacement happened at the top line, while maximum displacement occurred at the bottom line. In addition, the horizontal color intensity is uniform, while the vertical color intensity increases evenly from the top line to the bottom line. After applying the warping operation for the deformed image [Fig.~\ref{fig_ground_truth_raft}~(b)] using the computed deformation field [Fig.~\ref{fig_ground_truth_raft}~(d)], the corrected imaged is visualized in Fig.~\ref{fig_ground_truth_raft}~(c). The position of the anatomical structures on the corrected images is very close to the ground truth image obtained when the contact force is zero (a large amount of gel was filled between the probe and contact surface) in terms of the Normalized Cross Correlation (NCC) Coefficient (improved from $0.31$ to $0.99$), while the pixel intensity is more bright. Besides the synthetic data, we shown another representative example of a real deformed image obtained when the force is $7.5~N$ in Fig.~\ref{fig_ground_truth_raft}. The computed deformation field [Fig.~\ref{fig_ground_truth_raft}~(h)] is more complex, which represents individual pixel displacement direction and magnitude. After the warping process, the corrected image [Fig.~\ref{fig_ground_truth_raft}~(g)] also becomes very close to the ground truth (NCC similarity improved from $0.34$ to $0.86$).

\begin{figure}[h!]
    \centering
    \includegraphics[width=0.45\textwidth]{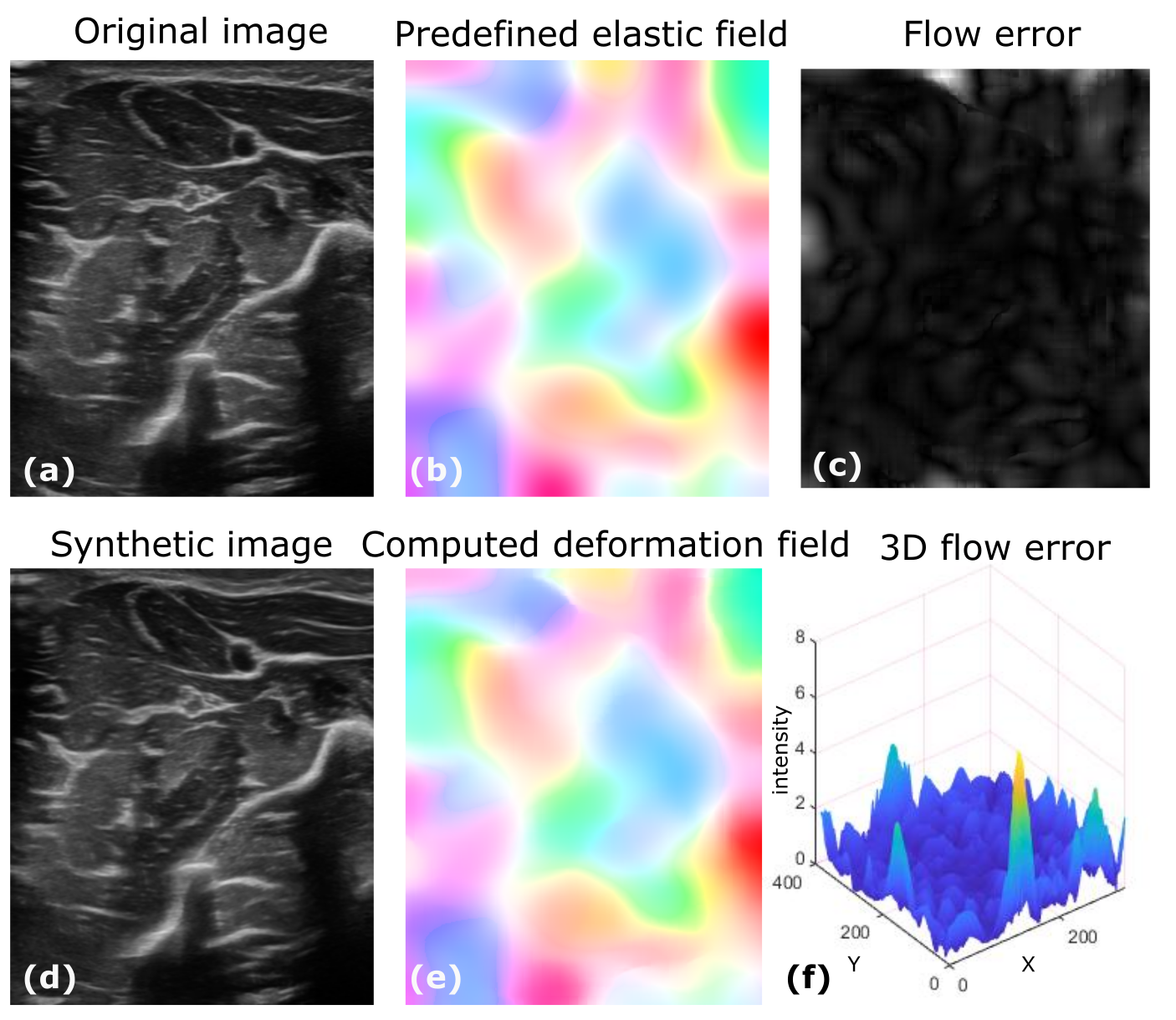}
    \caption{\revision{The performance of the estimation of a representative nonlinear elastic distortion field. (a) the original image. (b) the predetermined elastic deformation field using elastic distortion~\citep{simard2003best}. (d) the distorted image of (a) the original image by applying (b) the elastic deformation. (e) the computed deformation field using the modified RAFT. (c) and (f) are 2D and 3D visualization of the end-to-end point error (Euclidean distance) between (b) the predefined elastic field and (e) the computed deformation field.}
    }
    \label{fig_manual_flow}
\end{figure}

\par
Considering the US deformation is highly nonlinear due to the variations of tissue properties, we further validate the effectiveness of the modified RAFT for estimating non-linear deformation fields. To generate realistic distorted images, the elastic distortion~\citep{simard2003best} was randomly determined, and then the synthetic images can be obtained by applying the elastic distortion field to the original images. An intuitive example is demonstrated in Fig.~\ref{fig_manual_flow}. Based on the original and synthetic images, the deformation field can be calculated using RAFT [Fig.~\ref{fig_manual_flow}~(e)]. To quantitatively compare the known elastic field and computed deformation field, the end-to-end point error (Euclidean distance) is computed for individual pixel locations and visualized in the flow error map [see Fig.~\ref{fig_manual_flow}~(c)]. To make the flow error map more visible, the pixel intensity was considered as $[0,1]$ for visualization. The locations with large error values ($>1$ pixel) were considered as one in Fig.~\ref{fig_manual_flow}~(c). To reflect the error at each pixel location, the 3D error image is shown in Fig.~\ref{fig_manual_flow}~(f). The maximum distance error is $6.5$ pixels, while the mean error is only $0.5$ pixel. Furthermore, the NCC similarity value between the original and corrected images is enhanced to $0.98$ from $0.71$ between the original and synthetic images. Therefore, we consider the modified RAFT can provide decent deformation field ground truth for the development of learning-based deformation correction methods.

\subsection{Data Augmentation}
\par
To improve the diversity of datasets, some classic data augmentation methods, e.g., random scaling, rotation, and translation, are widely used in the field of computer vision. However, US images have different characteristics (e.g., physical sound attenuation and tissue-based distortion) from natural camera images. Regarding the force-induced deformed images, the displacements in the vertical and lateral directions are coupled. To keep this characteristic, only two augmentation methods were applied here. To reflect the anatomy-aware ability, B-mode images were horizontally flipped at random. Besides, a crop mask ($W_c\times H$, where $W_c\leq W$) was randomly initialized to obtain the cropped images, where the image height $H$ was kept as the original images. The reason for not changing $H$ is that the deformation is physically transmitted and accumulated from the skin surface. We consider these two augmentation methods can benefit to achieving anatomy-aware correction performance by diversifying the anatomy's location.


\section{Method}
\par
\begin{figure*}[h!]
    \centering
    \includegraphics[width=0.85\textwidth]{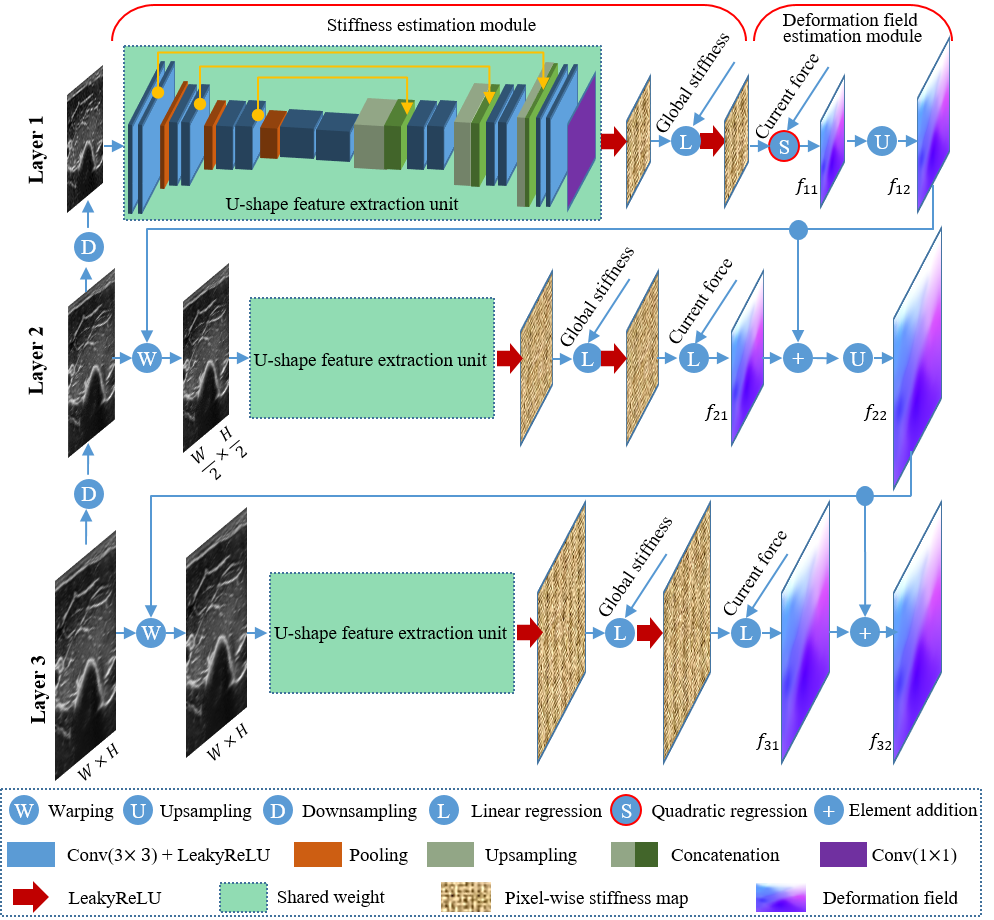}
    \caption{Illustration of the proposed DefCor-Net in a coarse-to-fine manner.
    }
    \label{fig_overview}
\end{figure*}

In this section, we present a novel CNN-based DefCor-Net organized in a coarse-to-fine manner for estimating the pixel-wise displacement maps of individual deformed US images acquired under various contact forces. By incorporating the concept of pixel-wise stiffness maps and patient-specific global stiffness obtained by robotic palpation, the proposed framework can predict the anatomy-aware deformation field. The proposed DefCor-Net consisting of three layers is depicted in Fig.~\ref{fig_overview}. To capture potential large deformation, the original B-mode images ($W\times H$) are down-sampled twice ($\frac{W}{2}\times\frac{H}{2}$ and $\frac{W}{4}\times\frac{H}{4}$) as the inputs of the first two layers. In each layer, the networks can be grouped into SEM and DFM modules, respectively. Detailed explanations of individual parts are provided in the following subsections.


\subsection{Pixel-Wise Stiffness Estimation Module}~\label{subsection: stiffness}
\par
Considering real scenarios, human tissues are non-homogeneous and vary largely from one patient to another one. Thereby, in order to estimate the deformation field only based on the deformed B-mode image without reference, it is important to compute the pixel-wise local stiffness. This means that the system needs to be aware of the anatomies' position on the resulting B-mode images. To this end, a U-shape feature extraction unit is initially employed to extract the initial stiffness map from a single deformed image. The initial stiffness map can indicate the pixel-wise nonlinear tissue properties. To further consider the patient-specific effects, a constant global stiffness is computed for each subject via robotic palpation and fed to a stiffness update module.

\subsubsection{U-Shape Feature Extraction}
\par
In order to effectively and robustly extract feature representations from images, Simonyan~\emph{et al.} proposed VGG-Net using very deep CNN structures~\citep{simonyan2014very}. Specific to B-mode images, Baumgartner~\emph{et al.} proposed a CNN-based SonoNet~\citep{baumgartner2017sononet}. The U-Net~\citep{ronneberger2015u} is another successful representative, and its variations have been widely used for segmenting detailed structures like vessel boundaries~\citep{jiang2021autonomous, prevost20183d}. In this work, we also use a U-shape network to extract the pixel-by-pixel stiffness representations from B-mode images. 

\par
The details of the used U-shape architecture are shown in Fig.~\ref{fig_overview}, which is similar to the 3D-Unet~\citep{cciccek20163d} by replacing the 3D operations (e.g., 3D convolutions and 3D pooling) with 2D operations. Considering the intermediate stiffness representations could be negative, the LeakyReLU function is employed to guarantee the backpropagation process of the network instead of ReLU activation. The size of the outputs is kept the same as the input images. Due to the coarse-to-fine manner, the U-shape feature extractor is reused three times in all three layers to generate pixel-wise stiffness cues. Based on the experimental results, the weight-sharing strategy is implemented to reduce the number of parameters while maintaining good performance. The image-based stiffness map $\mathbb{K}$ can be expressed as Eq.~(\ref{eq_local_stiffness}).

\begin{equation}~\label{eq_local_stiffness}
    \mathbb{K}_{us} = \text{LeakyReLU}\left(\text{UNET}(\mathbb{I}_{US})\right)
\end{equation}
where UNET means the U-shape feature extraction unit.

\subsubsection{Patient-Specific Stiffness Map Updating}
\par
\begin{figure}[ht!]
\centering
\includegraphics[width=0.45\textwidth]{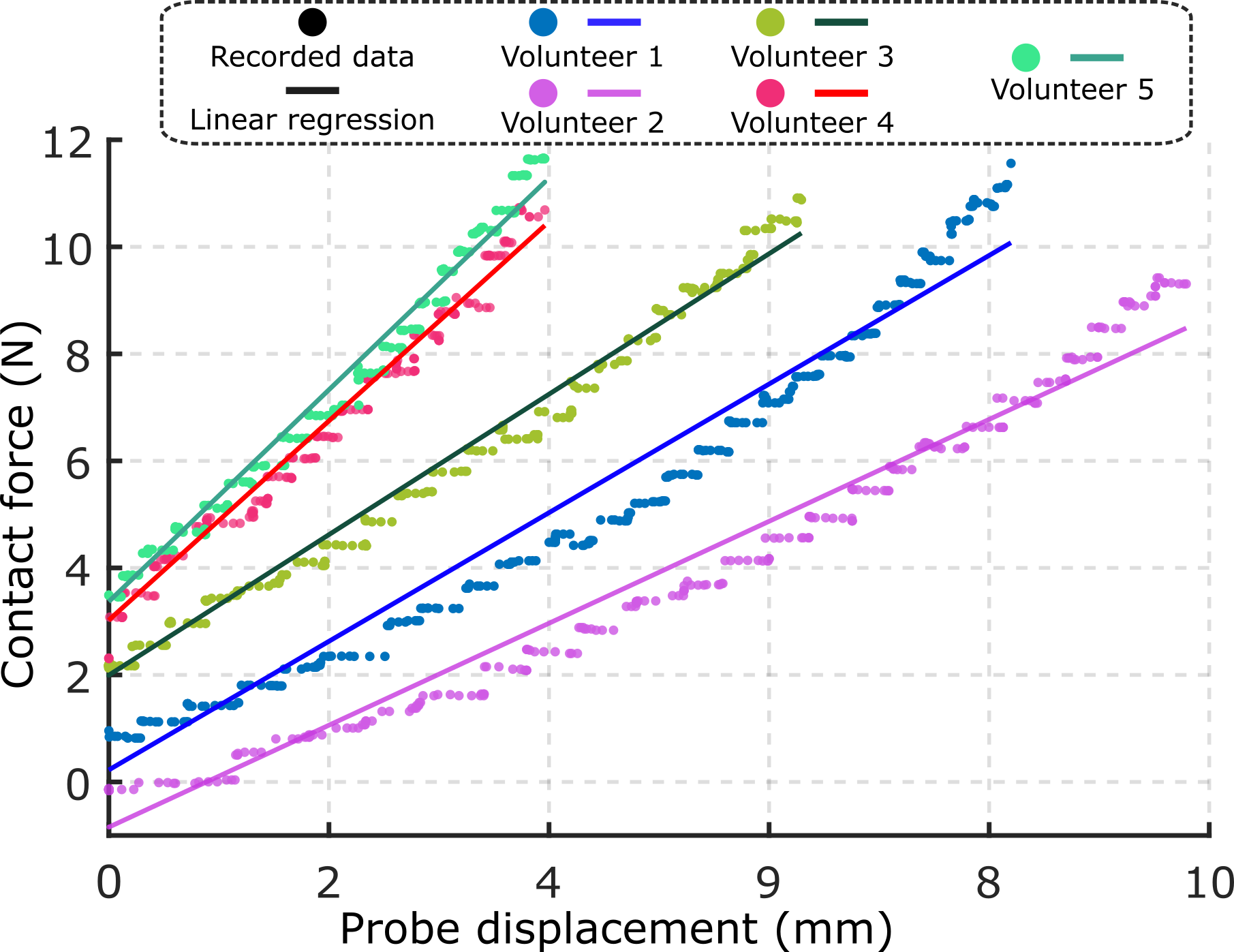}
\caption{\revision{Representative stiffness fitting results on the forearm of five volunteers.} The R-square \revision{values} of the optimized linear regressions are $0.96$, $0.97$, $0.99$, $0.99$, and $0.98$, respectively.
}
\label{Fig_volunteer_stiffness}
\end{figure}

In addition to the variation in stiffness between tissues of the same patient, the variation between patients also affects the accuracy of the stiffness representation. To consider the patient-specific factor in the proposed framework, a set of stiffness updating processes are further applied on the computed $\mathbb{K}_{us}$. To compute the global stiffness of individual patients, the robotic palpation was executed to record the real-time contact force in the direction of the probe centerline and the displacement of the probe tip in the force direction $\lambda_z$. Five representative data recorded from volunteers' forearms have been described in Fig.~\ref{Fig_volunteer_stiffness}. It can be seen from the recorded data that linear regression ($F_c = c_2\lambda_z + c_1$) is able to properly describe the relationship between the force and displacement. The R-square value of the computed regression models are $0.96$, $0.97$, $0.99$, $0.99$, and $0.98$, respectively. The high R-square fitness value means that the global stiffness $K_g$ for individual patients can be represented by the gradient $c_2$ of the optimized linear regression models. It is noteworthy that the computed global stiffness may not be desired for other parts of the same volunteer. It can only represent the tissue stiffness variable in the neighboring area of the sampling position.

\par
To stabilize the deviation of individual global stiffness $K_g$ for the same objects of different patients, $K_g$ of unseen patients is normalized using the Z-score approach before being used for adjusting the computed $\mathbb{K}_{us}$ as Eq.~(\ref{eq_nomolization}).

\begin{equation}~\label{eq_nomolization}
    k_g^n = \frac{k_g-\mu_g}{\delta_g}
\end{equation}
where $\mu_g=\sum_{i=1}^{N}\frac{1}{N}k_g^i$ and $\delta_g=\sqrt{\frac{\sum{k_g^i-\mu}^2}{N}}$ are the average and standard deviation of the recorded data, respectively. Then, the patient-specific stiffness map $\mathbb{K}_{us}^g$ is computed as Eq.~(\ref{eq_global_stiffness}).

\begin{equation}~\label{eq_global_stiffness}
   \mathbb{K}_{us}^g = \text{LeakyReLU}\left(c_1 \mathbb{K}_{us} + c_2 k_g^{n} + c_3\right)
\end{equation}

\subsection{Deformation Field Estimation Module}
\par
To recover a deformed image, a displacement field is required. To estimate the deformation field, polynomial regressions are often employed~\citep{sun2010trajectory, virga2018use, treece2002correction}. After comparing the recovery performance using different regression models, the quadratic model was found to give a better result than linear regression, while the performance was not improved significantly for a higher order polynomial model~\citep{sun2010trajectory}. Thereby, quadratic regressions were empirically used in the first layer to characterize the displacement field with respect to the current force $F_c$ and the pixel-wise stiffness map $\mathbb{K}_{us}^g$ as Eq.~(\ref{eq_sum_displacement}).

\begin{equation}~\label{eq_sum_displacement}
\left[\begin{array}{l}
\mathbf{D}_x(x,y) \\
\mathbf{D}_y(x,y)
\end{array}\right] =F_c \left[\begin{array}{l}
\mathbf{K}_x \\
\mathbf{K}_{y}
\end{array}\right] \mathbf{M}_{v}\left(\mathbb{K}_{us}^g(x, y)\right)
\end{equation}
where $(x, y)$ represents a pixel position on the image and $\mathbf{D}_x$ and $\mathbf{D}_y$ represent the pixel displacements in X (lateral) and Y (axial) directions, respectively. $\mathbf{M}_{v}(h)=\left[h^2, h, 1\right]^{T}$ ($h = \mathbb{K}_{us}^g(x, y)$), and $\mathbf{K}_{i} \in R^{1 \times 3}$ ($i=x$ or $y$) is the unknown parameters matrix of the quadratic regressions need to be optimized. To take the advantage of residual idea for the case with relative large deformation, the computed deformation field map $f_{11}$ is up-sampled as Eq.~(\ref{eq_upsampled}).

\begin{equation}~\label{eq_upsampled}
   f_{i2} = UP(f_{i1}, c_s)\revision{\otimes} c_s
\end{equation}
where $i=1$ or $2$ is the layer identification, $UP()$ is upsampling using bilinear interpolation, $c_s=2$ is the scaling coefficient, \revision{$\otimes$} is element-wise multiplication operation.  
 
\par
Compared with the methods that compute the deformation field only based on the contact force~\citep{sun2010trajectory, virga2018use}, the inclusion of tissue's mechanical properties theoretically allows the application of an optimized model for trained data on unseen objects by updating their specific stiffness. Preliminary work on this idea had been reported in~\citep{jiang2021deformation}, where the authors successfully applied an optimized regression model of a stiff phantom to a soft phantom with significantly different stiffness values. Although they considered the dynamic representation of the human tissue with respect to the applied force, the stiffness used there was considered homogeneous across all pixels in the image obtained under a specific force. This means anatomies' position changes on the deformed images (e.g., bone surface move from the left to the right side of the resulting images) cannot be properly addressed. To address this challenge, the pixel-wise stiffness map $\mathbb{K}_{us}^g$ is used in Eq.~(\ref{eq_sum_displacement}) instead of a single value to accurately estimate the deformation field, which is expected to be independent of the anatomy's position on deformed images.

\subsection{Coarse-to-Fine Deformation Field Extraction}
\par
In consideration of the possibility of relative large deformation, DefCor-Net is organized in a coarse-to-fine manner. After introducing the stiffness and deformation field estimation modules, this subsection elaborates on the contributions of individual layers and the whole workflow among the three layers. 

\par
Layer 1 is designed to deal with the overall large deformation using the downsampled inputs (size is $(\frac{W}{4}\times \frac{H}{4})$). Compared with the other two layers, a quadratic regress (Eq.~(\ref{eq_sum_displacement})) rather than a linear model is used in layer 1 to capture the non-linear properties of the deformation. After computing the deformation field in the first layer $f_{11}$, an up-sampling process based on the bilinear technique is implemented to obtain the field $f_{12}$ as the base deformation field of layer 2. To capture the deformation characteristics in the resolution of $(\frac{W}{2}\times \frac{H}{2})$, a warp operation is applied on the input images $\mathbb{I}_{in}$ as $warp(\mathbb{I}_{in}, f_{12})$. The warped results are further fed to the U-shape feature extraction unit with shared parameters to create an intermediate stiffness-related mask. Similar to layer 1, the patient-specific stiffness and the contact force $F_c$ are integrated consecutively using Eq.~(\ref{eq_global_stiffness}) and (\ref{eq_sum_displacement}) to compute the intermediate deformation field $f_{21}$. Based on the residual idea, the main deformation characteristics are accounted for in layer 1, while the residual pixel displacement is refined using images with higher resolutions in layers 2 and 3. Thus, a linear function ($\mathbf{M}_{v}^{'}(h)=\left[h, 1\right]^{T}$) is empirically used in Eq.~(\ref{eq_sum_displacement}) for the latter two layers. Then, the residual displacement field $f_{21}$ is added to the based flow $f_{12}$ using element-wise addition. Repeating the aforementioned steps in layer 3 and the final displacement field $f_{32}$ can be achieved. To train the network, the $f_{32}$ will be compared with the ground truth computed using RAFT.

%

\begin{table*}[!ht]
\centering
\caption{Summary of the Key Characteristics of Different Approaches}
\label{Table_key_features}
\resizebox{0.98\textwidth}{!}{
\begin{tabular}{ccccccc}
\noalign{\hrule height 1.2pt}
Approach & Axial def. & Lateral def. & Force cue    & Tissue property & Anatomy-aware  &In-vivo test \\ 
\noalign{\hrule height 1.0 pt}
Pheiffer~\emph{et al.}~\citep{pheiffer2015toward} & \cmark & \xmark & \xmark &\xmark &\xmark &\cmark \\ 
Dahmani~\emph{et al.}~\citep{dahmani2017model}    & \cmark & \cmark & \xmark &\cmark &\xmark &\xmark \\
Sun~\emph{et al.}~\citep{sun2010trajectory}      & \cmark & \cmark & \cmark &\xmark &\xmark &\xmark\\
Virga~\emph{et al.}~\citep{virga2018use}      & \cmark & \cmark & \cmark &\xmark &\xmark &\cmark\\
Burcher~\emph{et al.}~\citep{burcher2001deformation}    & \cmark & \cmark & \cmark &\cmark &\cmark &\xmark\\
Jiang~\emph{et al.}~\citep{jiang2021deformation} & \cmark & \cmark & \cmark &\xmark &\xmark &\xmark\\
Proposed DefCor-Net      & \cmark & \cmark & \cmark &\cmark &\cmark &\cmark\\

\noalign{\hrule height 1.2 pt}
\end{tabular}}
\end{table*}

\subsection{Loss Function}
\par
The proposed DefCor-Net is organized as a supervised learning problem. The ground truth of the deformation field $f_{gt}$ is computed as Section~\ref{sec:gt_raft}. The L1 loss is used to guide the learning process. Since the resolution of the input images are different, the sub-size $f_{gt}$ is computed as Eq.~(\ref{eq_downsample_gt}).

\begin{equation}~\label{eq_downsample_gt}
   f_{gt}^{\ell} = Down(f_{gt}, c_s^{2-\ell})~\revision{\otimes}~(c_s^{-1})^{2-\ell}
\end{equation}
where $\ell=1$ or $2$ is the layer iteration, $Down()$ is downsampling operation based on bilinear interpolation. The L1 loss of all three layers is computed as Eq.~(\ref{eq_L1}).

\begin{equation}~\label{eq_L1}
   \mathcal{L}_{l1}= \sum_{\ell} \left\|f_{gt}^{\ell}-f_{pred}^{\ell}\right\|_{1}
\end{equation}
where $f_{pred}=$\{$f_{11}$, $f_{21}$, $f_{32}$\} is the predicted deformation field.  

\par
Besides $\mathcal{L}_{l1}$, the geometry continuity of the displayed anatomies should be further considered, which is important for clinical diagnosis, in particular for the future development of advanced computer-assisted diagnosis. To encourage such behavior, a smoothness loss $\mathcal{L}_{sm}$ proposed by~\citep{jonschkowski2020matters} is modified as Eq.~(\ref{eq_smooth_loss}).

\begin{equation}~\label{eq_smooth_loss}
\begin{split}
\mathcal{L}_{sm}=\sum_{k} \exp \left(-\lambda_x^k \left(\frac{\partial I}{\partial x}\right)^2\right ) f(\frac{\partial^{k} f_{32}}{\partial x^{k}})\\
+ \underbrace{\exp \left(-\lambda_y^k \left(\frac{\partial I}{\partial y}\right)^2\right)}_{geometry} \underbrace{f(\frac{\partial^{k} f_{32}}{\partial y^{k}})}_{smoothness}
\end{split}
\end{equation}
where $k={1, 2}$ represents the first- and second-order term. The smoothness term mainly consists of the partial derivatives of predicted deformation map $f_{32}$ with respect to $x$ and $y$, respectively. By minimizing the derivatives, the intensity of $f_{32}$ becomes smooth. To avoid zero gradient, f(m)= $\sqrt{m^2+\epsilon^2}$. However, the smooth term will decrease anatomies' geometry accuracy. To address this, the geometry term is used as edge coefficient. The boundary can be identified by computing the gradient image of the original inputs with respect to $x$ and $y$, respectively. The geometry term is used to ensure the boundaries of different anatomies are not smoothed out.  

\par
Then, the final training loss is organized as Eq.~(\ref{eq_Loss_all}).

\begin{equation}~\label{eq_Loss_all}
   \mathcal{L}=\lambda_{1} \mathcal{L}_{l1}+\lambda_{2} \mathcal{L}_{sm}
\end{equation}
where $\lambda_{1}$ and $\lambda_{2}$ are the hyperparameters to combine the two different loss. Based on the output performance, $\lambda_{1}$ and $\lambda_{2}$ are empirically set to $1$ and $10$, respectively.

\section{Results}
\par
To highlight the contributions of the proposed approach, we provide a summary of the key characteristics of existing methods (see TABLE~\ref{Table_key_features}) that aim to correct force-induced US deformation. In addition to the metric of axial and lateral deformation correction ability, the terms of force cue and tissue property indicate whether the contact force and tissue mechanical property are used for US image deformation correction. The approaches without considering tissue properties will limit their performance in generalization capability among patients in theory. The metric of anatomy-aware ability indicates whether the method can recover a deformed image correctly regardless of the relative position of the anatomy of interest in US images, such as a bone surface located on the left or right side in the deformed image.
The anatomy-aware ability is important in real scenarios because the position of target tissues varies significantly according to the placement of the probe. Most of the previous work cannot achieve anatomy-aware performance because they usually directly characterize the deformation with respect to the applied force using polynomial regression models. Due to the absence of tissue properties, the optimized regression model is the overall optimal interpreter for seen data, while its inference does not account for the positional variation of anatomy in US images.
FEM-based approaches~\citep{burcher2001deformation} could be one of the potential solutions. Nevertheless, the need for precise estimation of patient-specific tissue parameters remains an open challenge in the community. In addition, FEM is time-consuming and cannot be used in real-time based on current developments.

\par
To address this challenge, we use advanced deep learning to implicitly estimate the pixel-wised stiffness characteristics form input US images. By forcing DefCor-Net to explicitly infer nonlinear tissue stiffness maps, the presented supervised learning framework can compute the deformation field based on both applied force and pixel-wise stiffness maps. Consequently, anatomy-aware performance is implicitly achieved by being aware of the changes in stiffness value distributions in the inferred stiffness maps.Due to the fact that the stiffness value changes should be continuous, the intermediate variables (i.e., stiffness maps) are qualitatively validated whether being consistent with the underlying physics in Section-\ref{sec:stiffness_vali}. In addition, quantitative evaluations of the proposed DefCor-Net are performed on in-vivo limb US images. The validation of the anatomy-aware ability has been demonstrated in Section-\ref{sec_vali_anatomy_aware}. The quantitative evaluation of the proposed deformation correction method based on local bone is summarized in Section~\ref{sec_vali_quantitative}. Finally, the target localization accuracy and dense flow error over the entire images are computed and discussed in Section~\ref{sec_vali_quantitative_overall}.

\begin{figure*}[h!]
    \centering
    \includegraphics[width=0.80\textwidth]{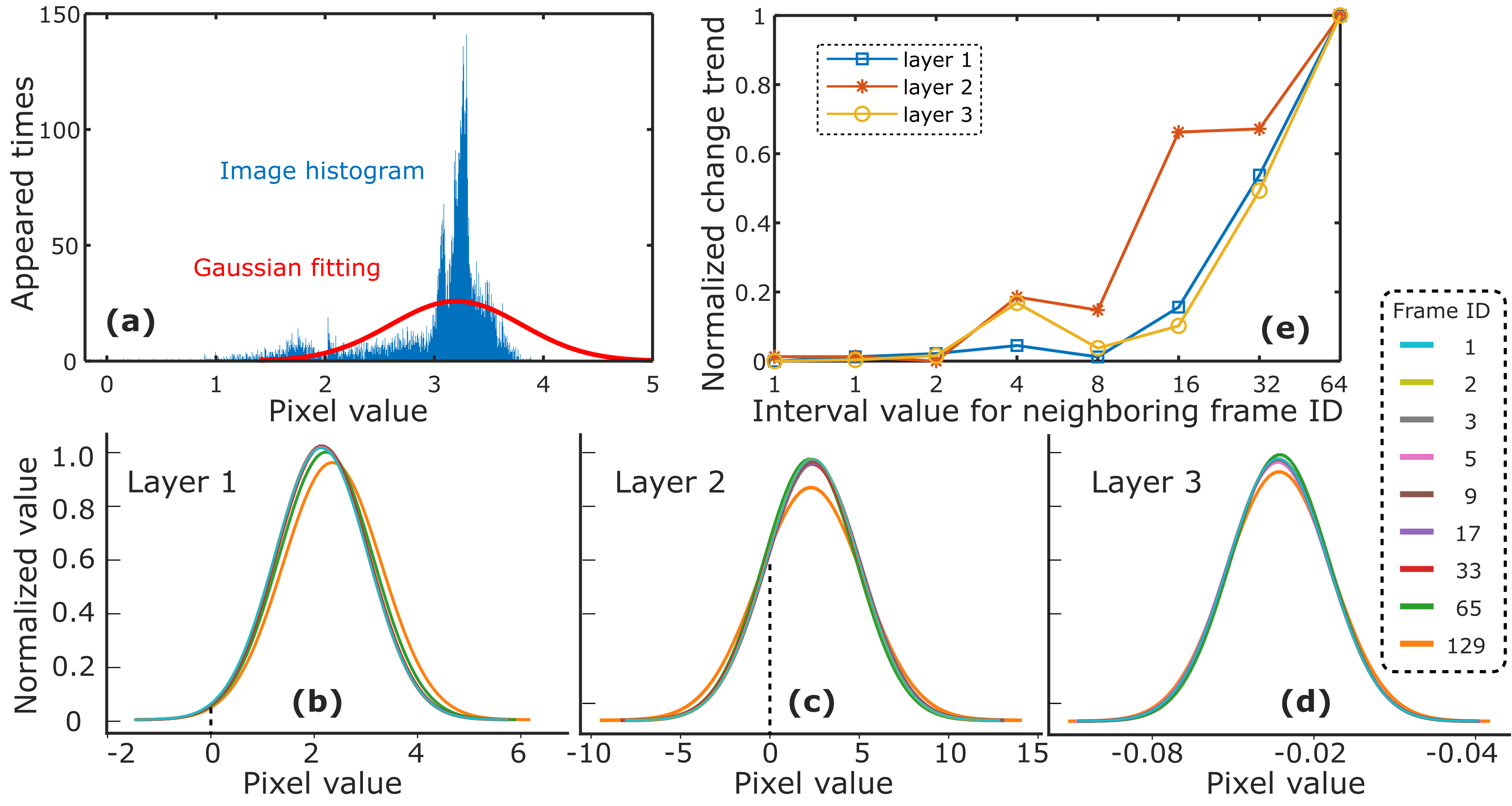}
    \caption{Consistency of intermediate stiffness inferences with respect to the similarity levels of input images. (a) Representative images histogram and the fitted Gaussian distribution of an inferred stiffness map in layer 1. (b), (c), (d) are the fitted Gaussian distributions of the selected frames in layers 1, 2, and 3, respectively. (e) the normalized difference of the exponentially increased intervals of neighboring frames.    
    }
    \label{fig_guassian_change}
\end{figure*}

\subsection{Implementation Details}
\par
The coarse-to-fine DefCor-Net was implemented using PyTorch framework. The training and evaluation configurations were maintained based on OpenMMLab\footnote{https://github.com/open-mmlab}. The DefCor-Net model was optimized using the ADAM optimizer~\citep{kingma2014adam} on a workstation with a GPU of GeForce GTX 1080. The learning rate was $0.0001$ for $50k$ epochs. To guarantee the generability of the trained model for different patients, a small batch size (four) was used in this work. 
The sweeps used for testing are not used in the training data set. The image size is $320\times384$. To improve the diversity of the training dataset for achieving anatomy-aware effects, the images were randomly cropped from the original size to $256\times 384$. The training process took around eight hours. In addition, the processing time (Mean$\pm$SD) for generating the deformation map was $34\pm10~ms$ over $844$ samples, which means that the approach is able to be used for the application expecting real-time performance.


\subsection{Consistency of Intermediate Stiffness Inferences}~\label{sec:stiffness_vali}
\par
The proposed DefCor-Net is designed by implicitly considering biomedical knowledge. It can be seen from Fig.~\ref{fig_overview} that the inferred pixel-wise stiffness map is an important intermediate variable. Since it is not possible to obtain the ground truth of pixel-wised stiffness of the input B-mode images, we cannot directly compare the inferred intermediate stiffness-related terms. However, stiffness is a physical (analog) variable, which has the characteristic that the changes in stiffness are continuous in the ideal case.

\par
To validate the consistency among the automatically generated intermediate stiffness maps, a set of unseen images from the same US sweep were fed to the well-trained DefCor-Net. The inferred stiffness maps (the latter ones) in three layers were stored. To assess the difference between different images, the results of nine representative frames have been shown in Fig.~\ref{fig_guassian_change}. The frame ID are exponentially increased ($1$, $1+2^{0}$, $1+2^{1}$, $\cdots$, $1+2^{7}$). To quantitatively assess the consistency for neighboring sampled frames, the image histogram of all obtained stiffness maps is drawn, where the number of bins was fixed to $1000$ in this work. Then, the Gaussian fitting was repeatedly carried out to fit all histograms. Representative historiography and the corresponding Gaussian fitting result of the inferred stiffness map obtained in layer 1 are shown in Fig.~\ref{fig_guassian_change}~(a). All $27$ fitted Gaussian distributions of layer 1, layer 2 and layer 3 are depicted in Figs.~\ref{fig_guassian_change}~(b), (c) and (d), respectively. To intuitively display the difference between the selected nine frames, we computed the overlap area $A_{ov}$ between the fitted Gaussian distributions obtained from two neighboring sampled frames. Since the cumulative distribution function for a normal distribution is always one when $x=+\infty$, the difference between two fitted Gaussian distributions is represented by $1-A_{ov}$. The normalized changes between neighboring frames for all three layers are depicted in Fig.~\ref{fig_guassian_change}~(e). 

\begin{figure*}[h!]
    \centering
    \includegraphics[width=0.95\textwidth]{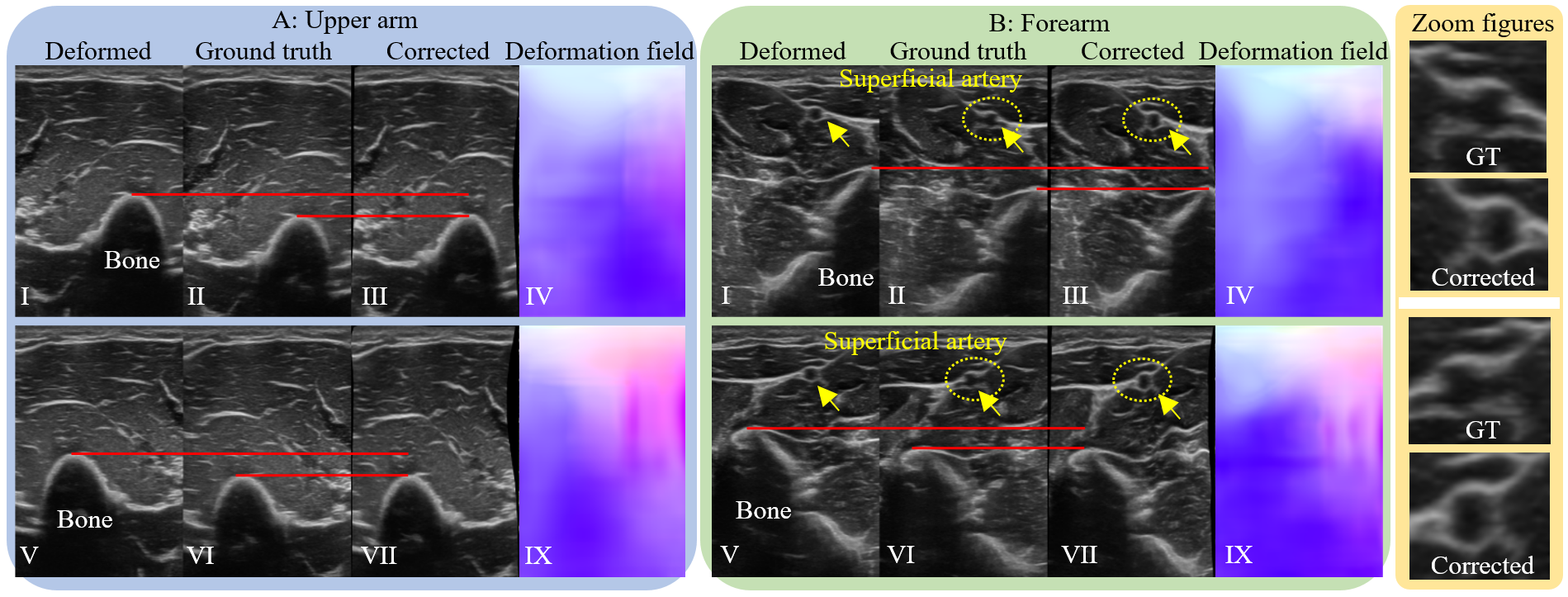}
    \caption{Illustration of the anatomy-aware characteristic. The upper arm and forearm US images are recorded when the force is $7.0~N$ and $6.9~N$, respectively. The superficial artery on uncompressed and corrected images is zoomed in and depicted in the right column.
    }
    \label{fig_anatomy_aware}
\end{figure*}

\par
Due to the coarse-to-fine structure, the non-linear stiffness properties are mainly modeled in the first layer (quadratic regression). Thus, the inferred stiffness map of layer 1 is corresponding to the real stiffness, while the ones of the other two layers are considered residual compensations. It can be seen from Figs.~\ref{fig_guassian_change}~(a) and (b) that the pixel values of the inferred stiffness map from layer 1 are mainly distributed around two, which is consistent with the prior knowledge that the stiffness term is positive ideally. It is noteworthy that there is no such restriction on the residual compensations obtained in the other two layers. The exponentially increased intervals between neighboring frames are used to represent the different levels of similarity between the input B-mode images. Due to the characteristic of the analog variable, the stiffness changes should be positively corresponding to the similarity of the input images. This phenomenon is demonstrated in Fig.~\ref{fig_guassian_change}~(e). The changes are very small ($<0.02$ when the frame interval is small than two) for the automatically inferred stiffness maps in all three layers. Such changes become larger when the interval increases. There is an exceptional case when the frame interval is four, where the changes are larger than the results obtained when the interval increased to eight (layer 1: $0.04$ vs $0.01$; layer 2: $0.18$ vs $0.15$; layer 3: $0.16$ vs $0.04$). But regarding the stiffness term (layer 1), $0.04$ still can be seen as no significant difference between the neighboring frames. The experimental results demonstrate that the difference between the inferred stiffness maps is consistent with the similarity levels of the input images.

\subsection{Validation of Anatomy-Aware Characteristic}~\label{sec_vali_anatomy_aware}
\par
To validate the performance of the advanced anatomy-aware ability, experiments were carried out on unseen sweeps on both the upper arm and forearm. Since the stiffness of the bone structure is far higher than the surrounding soft tissues, the bone structure will be preserved during the robotic palpation. Thereby, the pixel position of the bone on the images was used to intuitively demonstrate the changes caused by the deformation and the quality of the corrected results in Fig.~\ref{fig_anatomy_aware}. For each case (upper arm or forearm), two representative deformed images with significantly different anatomy (bone) positions were used to demonstrate the anatomy-aware ability of the proposed DefCor-Net. 

\par
It can be seen from Fig.~\ref{fig_anatomy_aware} that the trained DefCor-Net can properly recover the deformed images in all four representative cases for different tissues. The deformed images are significantly different from the ground truth, while the corrected ones become very close to the ground truth. As shown in Fig.~\ref{fig_anatomy_aware}, the position of dark blue parts on the deformed fields is consistent with the positions of the bones on their corresponding inputs. Since the bone structure is not deformed, the displacement of the pixels covered by the bone surface will be consistent in both direction and amplitude. This phenomenon has been demonstrated by the quasi-homogeneous intensity values (dark blue area) on the computed deformation field. In addition, it is noteworthy that the superficial artery geometry on the corrected images is more clear than the one displayed on the ground truth. This phenomenon is caused due to the inherited characteristic of US modality, which requires a certain force to achieve optimal acoustic coupling performance. To avoid imaging deformation, the ground truth is obtained by applying zero pressure ideally. This results in the less clear boundary of the superficial artery on the shallow layer of the ground truth than of corrected images. Two representative comparisons are demonstrated in the far right column in Fig.~\ref{fig_anatomy_aware}. The vascular boundaries on corrected images are more clear and more complete than the ones on the natively uncompressed images.

\begin{table*}[!ht]
\centering
\caption{Performance of the Corrected Results using Different Approaches \revision{(Dice Coefficient)}}
\label{Table_dice_score_bone}
\begin{tabular}{cccccc}
\noalign{\hrule height 1.2pt}
Contact Pressure & Deformed image &Linear Scaling &  DefCor-Net-C & DefCor-Net-Lin  & DefCor-Net \\ 
\noalign{\hrule height 1.0 pt}

1 N & $84.9\pm15.5$    & $86.6\pm15.1$    &  $89.0\pm12.8$     & $90.2\pm9.4$   & \textbf{95.9$\pm$3.3} \\ 
2 N & $61.0\pm28.2$    & $76.2\pm14.8$    &  $78.6\pm19.8$     & $88.6\pm4.4$  & \textbf{92.4$\pm$4.8}  \\ 
3 N & $43.2\pm29.1$    & $71.5\pm10.0$    &  $55.7\pm25.6$     & $85.4\pm6.6$  & \textbf{91.1$\pm$7.5} \\ 
4 N & $30.4\pm30.2$    & $74.4\pm18.0$    &  $47.6\pm28.5$     & \textbf{87.8$\pm$6.8}  & 87.8$\pm$7.1 \\ 
5 N & $19.5\pm25.1$    & $65.3\pm15.6$    &  $40.3\pm31.8$     & $84.1\pm11.5$  & \textbf{87.8$\pm$12.2} \\ 
6 N & $14.3\pm20.9$    & $57.2\pm22.8$    &  $32.1\pm28.9$     & $81.5\pm7.7$   & \textbf{82.6$\pm$12.1} \\
\noalign{\hrule height 1.2 pt}
\end{tabular}
\end{table*}

\subsection{Evaluation of Deformation Correction on Bone structure}~\label{sec_vali_quantitative}
\subsubsection{Bone Segmentation}~\label{sec:bone_seg}
\begin{figure}[h!]
    \centering
    \includegraphics[width=0.45\textwidth]{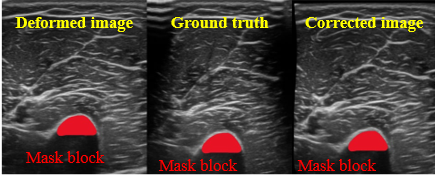}
    \caption{Illustration of a manually annotated mask block on deformed images, ground truth, and corrected images. The deformed image was recorded on a volunteer's upper arm when the contact force was $4~N$. 
    }
    \label{fig_human_label}
\end{figure}

Due to the large stiffness against the surrounding soft tissues, the bone structure will not be deformed during scans. Taking advantage of this characteristic, we used the same mask block to partially annotate the bone structure on the deformed images, correct images, and ground truth from the whole palpation sweep (see Fig.~\ref{fig_human_label}). The top outline of the block needs to be consistent with the bone surface, while the bottom outline of the block was manually determined. Compared with the approach to annotating all images individually, a generic mask block can maximally minimize the inconsistency introduced by manual annotations. The use of a block area rather than the bone outline targets the same issue. A representative annotated result on a set of deformed, corrected images and ground truth is depicted in Fig.~\ref{fig_human_label}.

\subsubsection{Quantitative Evaluation}
\par
To provide quantitative assessment, the popular metric of \revision{Dice Coefficient} $d_c = 100\times\frac{2~|G\cap S|}{|G|+|S|}$ was computed between the ground truth and deformed/corrected images. $G$ represents the number of pixels in the mask block of the ground truth image, whereas $S$ is the number of pixels in the mask block of the deformed/corrected images. To demonstrate the superiority of the proposed DefCor-Net, it was compared with other approaches, including a linear deformed correction approach and various modifications of DefCor-Net. To fully validate the correction performance on the cases with different levels of deformation, the deformed images recorded with different forces ($[1,6]$) were sampled. To count the variations between different tissues (with significantly different stiffness), seven and three sweeps from the upper arm and forearm, respectively, were used for the validation. The detailed results are depicted in TABLE~\ref{Table_dice_score_bone}.

\par
Besides the Dice Coefficient computed between the ground truth and the deformed images, TABLE~\ref{Table_dice_score_bone} also presents the Dice Coefficient computed between the ground truth and the corrected images obtained by various approaches. The linear scaling approach assumes that the compression is linearly increased as the image depth. The DefCor-Net-C represents the network replacing the U-shape feature extraction unit with a classic 2D U-Net~\citep{ronneberger2015u}. The U-shape unit used in the final DefCor-Net is inspired by the 3D U-Net~\citep{cciccek20163d}, which has three layers with fewer parameters compared with the classic U-Net. The numbers of parameters in DefCor-Net-C and DefCor-Net are around $31$ million and $8$ million, respectively. The DefCor-Net-Lin represents the network replacing the quadratic regression with a linear regression in the first layer.

\par
It can be seen from TABLE~\ref{Table_dice_score_bone} that the Dice Coefficient of the recorded deformation images is dramatically decreased from $84$ to $14$ when the contact force increased from $1~N$ to $6~N$. This means that the bone's position on the US images is significantly changed when the contact force is increased. Therefore, to provide accurate anatomical geometry and position for diagnosis, it is necessary to apply a certain deformation correction. Based on TABLE~\ref{Table_dice_score_bone}, the highest Dice Coefficient is achieved when the force is $1~N$ for all individual cases. This is because the deformation is tiny and relatively easy to be recovered when the force is small ($1~N$). Regarding the case with a large force, the Dice Coefficients achieved by DefCor-Net are much higher than the ones achieved by linear and DefCor-Net-C ($82.6$ vs $32.1$ when force is $6~N$). Besides, it is noteworthy that DefCor-Net-Lin achieved the closest results to DefCor-Net, while still worse than DefCor-Net in most cases. This is because the deformation is not linearly with respect to the applied force and tissue stiffness. This is also consistent with the results achieved by~\citep{sun2010trajectory} that a high-order polynomial model outperforms the linear model for correcting force-induced deformation. The results demonstrate that the proposed DefCor-Net can effectively recover the deformed images.

\subsection{Evaluation of Deformation Correction on Overall Images}~\label{sec_vali_quantitative_overall}
Due to the bone structure often located in the lower part of images, it can only reflect the performance of the correction method over local areas. To comprehensively validate the performance of the proposed method, we further calculated the target localization accuracy (TLA) based on biomedical interfaces of different tissues (e.g., fat and muscle) on US images. Besides, a direct comparison was carried out between the computed deformation field of DefCor-Net and the ground truth calculated by RAFT. These two validations can intuitively demonstrate the performance of the deformation correction method over the whole view of B-mode images.

\begin{figure}[h!]
    \centering
    \includegraphics[width=0.25\textwidth]{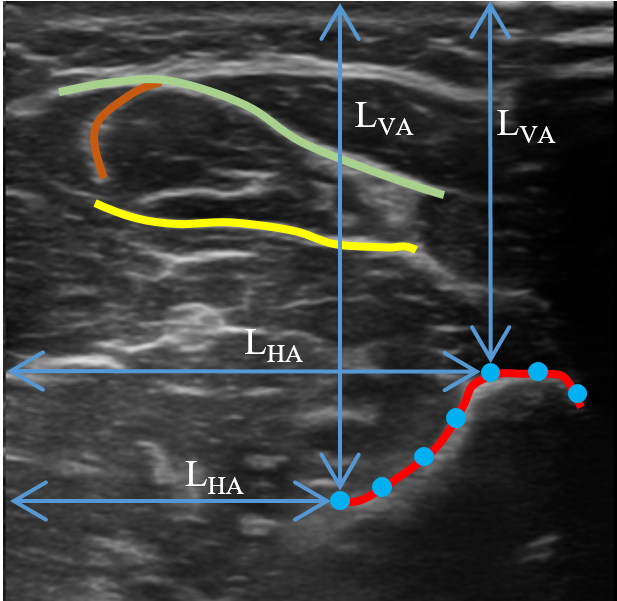}
    \caption{\revision{Illustration of tissue interface annotations on B-mode images. The four curves with clear ending points on both sides are carefully annotated. The blue dots are the sampling points generated on one selected interface. }
    }
    \label{fig_target_error}
\end{figure}

\subsubsection{Target Localization Accuracy}~\label{sec:Target_Localization_Error}
\par
US images often exhibit distinct curves caused by the boundaries between various tissues. Such interfaces located at different parts of the image can be used as good biomarkers to calculate the TLA between the uncompressed images and deformed/corrected images. A representative illustration is depicted in Fig.~\ref{fig_target_error}, where we carefully annotated the selected four distinct interfaces. Each selected interface should have clear ending points at both sides to ensure its completeness on a set of corresponding uncompressed, deformed, and corrected images. To compute TLA, we first evenly (every $5$ pixels in the horizontal axis) sampled the measurement points from all interfaces. Then horizontal and vertical distances ($L_{HA}$ and $L_{VA}$) at each sampling point were computed (see Fig.~\ref{fig_target_error}). Then TLA at each sampling point can be calculated as $HA = \frac{\|L^{i}_{HA}-L^{gt}_{HA}\|}{L^{gt}_{HA}}$ and $VA = \frac{\|L^{i}_{VA}-L^{gt}_{VA}\|}{L^{gt}_{VA}}$, where $i$ is the deformed or corrected images. The average TLA (HA and VA) computed on unseen scans from six volunteers are summarized in TABLE~\ref{tab:target_errro}.

\par
It can be seen from TABLE~\ref{tab:target_errro} that both HA and VA decrease when the contact force increases. The HA decreases from $0.98\pm0.04$ to $0.77\pm0.20$, while the VA decreases from $0.98\pm0.01$ to $0.84\pm0.09$. The TLAs are significantly enhanced after applying the proposed deformation correction process. Both HA and VA are maintained over $0.9$ in different force levels. The HA and VA are increased from $0.77\pm0.20$ to $0.91\pm0.08$ and from $0.84\pm0.09$ to $0.95\pm0.03$, respectively, when the force is $6~N$.  It is also noteworthy that SD is significantly reduced after deformation correction in most cases. This implicitly demonstrates the corrected performance is stable and robust among all sampling points on the images.

\begin{table}[tb]
\centering
\caption{\revision{Results in Terms of Target Localization Accuracy [Mean(SD)]}}
\label{tab:target_errro}
\resizebox{0.48\textwidth}{!}{
\revision{
\begin{tabular}{cccccc}
\noalign{\hrule height 1.2 pt}
\multirow{2}{*}{Pressure} & \multicolumn{2}{c}{ Horizontal accuracy (HA) } & & \multicolumn{2}{c}{ Vertical accuracy (VA) } \\
\cline { 2 - 3 } \cline { 5 - 6 } & Deformed & Corrected & & Deformed & Corrected  \\
\hline 
1 N & 0.95±0.04 & 0.96±0.03 & & 0.98±0.01 & 0.98±0.01 \\
2 N & 0.89±0.09 & 0.96±0.03 & & 0.94±0.07 & 0.97±0.03 \\
3 N & 0.85±0.13 & 0.94±0.04 & & 0.91±0.07 & 0.97±0.03 \\
4 N & 0.80±0.15 & 0.91±0.08 & & 0.88±0.09 & 0.95±0.04 \\
5 N & 0.79±0.16 & 0.90±0.08 & & 0.85±0.11 & 0.93±0.07 \\
6 N & 0.77±0.20 & 0.91±0.08 & & 0.84±0.09 & 0.95±0.03 \\
\noalign{\hrule height 1.2 pt}
\end{tabular}
}}
\end{table}

\subsubsection{Dense Error Map of Deformation Field}~\label{sec:Flow_Error_Comparison}
\par
\begin{figure}[h!]
    \centering
    \includegraphics[width=0.35\textwidth]{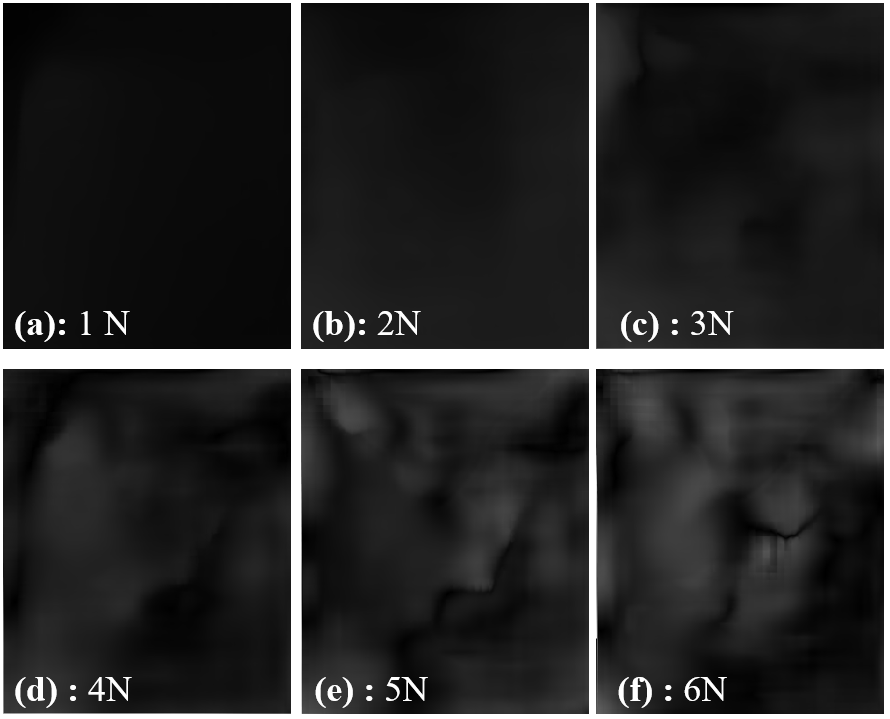}
    \caption{\revision{Illustration of the error maps between the estimated deformation field and the ground truth flow computed using RAFT. }
    }
    \label{fig_flow_error_map}
\end{figure}

To further provide dense error maps, we directly computed the end-to-end point error (Euclidean distance) between the DefCor-Net-calculated deformation field and the RAFT-calculated flow maps. Fig.~\ref{fig_flow_error_map} depicts representative dense error maps obtained under varying forces. To enhance visibility, the intensity values in all error maps were scaled by the same factor ($15$). The average pixel-by-pixel errors are $0.88\pm0.21$ pixels, $1.80\pm0.46$ pixels, $1.87\pm0.71$ pixels, $1.92\pm1.02$ pixels, $2.47\pm1.18$ pixels and $2.73\pm1.59$ pixels when the forces are $1~N$, $2~N$, $3~N$, $4~N$, $5~N$ and $6~N$, respectively. To ensure the deformation correction performance is robust to the pixel locations, we further counted the number of pixels (percentage $Per_{10}$, $Per_{15}$ and $Per_{20}$) with the error larger than $10$ pixels ($1.1~mm$), $15$ pixels ($1.76~mm$) and $20$ pixels ($2.34~mm$). Regarding Figs.~\ref{fig_flow_error_map} (a), (b) and (c), all errors are less than $10$ pixels. In Figs.~\ref{fig_flow_error_map} (e), (f) and (g), $Per_{10}$ increases to $2.5\%$, $2.0\%$ and $2.2\%$, while $Per_{20}$ are only 0, $0.6\%$ and $0.1\%$, respectively. This indicates that the computed deformation field using the proposed DefCor-Net is close to the ground truth flow. Therefore, we consider the proposed pixel-wise deformation correction can properly recover deformed US images caused by external pressure.

\section{Discussion}
\par
The proposed novel DefCor-Net works boldly in our setup. By explicitly incorporating pixel-wise stiffness information into the loop, the estimated deformation field can be generated implicitly with respect to the location of the anatomies in real time. But, there are some limitations of the current approach that are worth noting. First, this work only aims to correct the force-induced deformation, where the contact force is the main source for the resulting deformation in images. There is still a gap in propagating the proposed method to internal organs such as the liver, in which the deformation is a result of the coupling between external pressure and physiological motions such as respiration.
Second, this study only considered the data recorded from healthy volunteers. Although it works boldly in our setup, the applicability to patients with abnormal tissues, such as peripheral arterial disease, has not been validated yet. In the future, further studies should include patients' data to prove clinical transferability.

\section{Conclusion}
\par
This work first proposes a learning-based approach DerCor-Net for achieving accurate anatomical images from deformed B-mode images caused by the inevitable pressure. The novel DerCor-Net is organized in a coarse-to-fine fashion and incorporates the biomedical knowledge between force, stiffness, and contact force, which theoretically enables the extrapolation of applying a well-trained model to different patients. Benefited from the deep neural network, over $8$ million trainable parameters allow DerCor-Net to compute pixel-wise stiffness based on the input images and the deformation field ground truth obtained using an optical flow network RAFT~\citep{teed2020raft}. Besides improving the correcting accuracy, the online estimated pixel-wise stiffness map also enables the anatomy-aware characteristics for recovering the deformed image. This feature is still missed in the previous related work~\citep{sun2010trajectory, dahmani2017model, jiang2021deformation, virga2018use}, however, is very important in real scenarios where the sonographer cannot always maintain the target anatomy at the same position on the US view. To validate the proposed approach, DefCor-Net was trained using the data recorded from six volunteers, both forearms and upper arms, and further validated on $13$ unseen sweeps. The proposed DefCor-Net can significantly improve the accuracy of tissue geometry from deformed images (Dice Coefficient: $82.6\pm12.1$ vs $14.3\pm20.9$ when force is $6~N$). The average Dice Coefficient of the proposed method for the images with different levels of deformation achieves $89.6$. 

\par
We hope this work can reduce intra- and inter-operator variations of US acquisition by compensating for the force-induced deformations, thereby resulting in consistent and accurate diagnosis in clinical practice. This is particularly important for the further development of advanced computer-assisted diagnosis programs because the sensing and perception ability of machines are still significantly inferior to those of human operators. The future work includes integrating the proposed method to improve image-guided orthopedic surgery and exploring the way to extend this work in 3D by considering 3D continuity.

\appendix
\section{Optical Flow Color Coding}~\label{sec_append_color}


\begin{figure}[h!]
    \centering
    \includegraphics[width=0.25\textwidth]{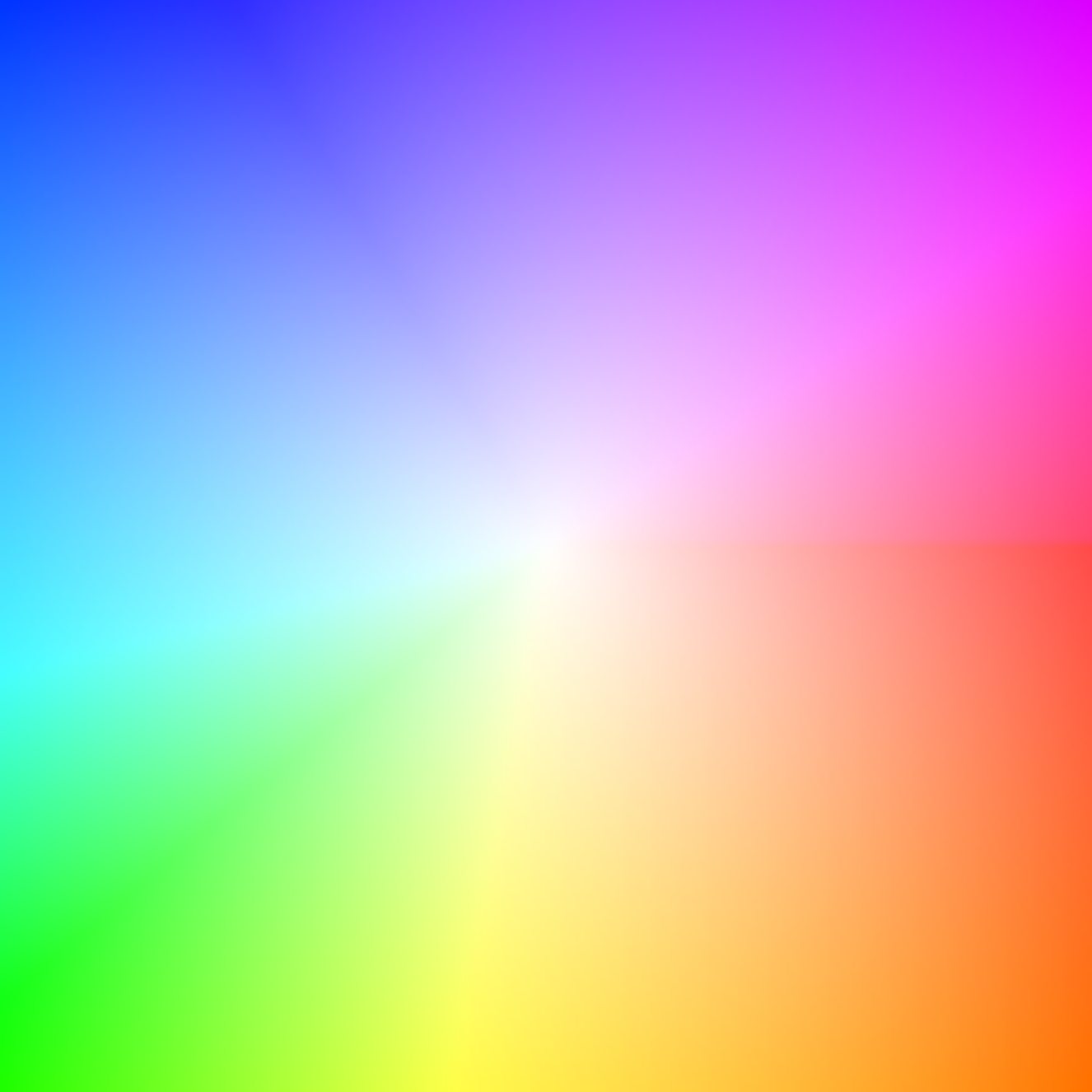}
    \caption{Illustration of the flow color encoding used in this article.}
    \label{fig_color_encoding}
\end{figure}

\par
In order to properly visualize computed optical flows, Baker~\emph{et al.} applied different colors to represent pixel displacement vectors~\citep{baker2011database}. The color coding is shown in Fig.~\ref{fig_color_encoding}, in which the hue represents the flow directions and the intensity indicates the relative magnitude. The displacement of each pixel in the figure is the vector from the square center to this pixel, while the center pixel (white color) does not move.



\section*{Declaration of Competing Interest}
\label{competinginterest}

The authors report no conflicts of interest.


\section*{Acknowledgments}
\label{acknowledgments}
\revision{The authors would like to acknowledge the Editors and anonymous reviewers for their time, and implicit contributions to the improvement of the article's thoroughness, readability, and clarity.}


\bibliographystyle{model2-names.bst}\biboptions{authoryear}
\bibliography{references}



\end{document}